\documentclass[conference]{IEEEtran}
\usepackage{graphicx}
\usepackage{amsmath}
\usepackage{newtxmath}
\usepackage[nobreak]{cite}
\usepackage{framed}
\usepackage{multirow}
\usepackage{xspace}
\usepackage{url}

\usepackage{cleveref}
\Crefname{figure}{Figure}{Figures}
\crefname{figure}{Fig.}{Figs.}
\crefname{table}{Table}{Tables}
\crefname{section}{Section}{Sections}
\Crefname{section}{Section}{Sections}

\newcommand{\Heading}[1]{\textbf{#1.}}
\newcommand{\RQ}[1]{\textit{RQ}${}_{\mathrm{#1}}$}
\usepackage{framed}
\newcommand{\Conclusion}[1]{\begin{framed}\noindent #1\end{framed}}

\def\etal{\textit{et al.}\xspace}
\def\ie{\textit{i.e.}}

\def\tf{\mathit{tf}}
\def\idf{\mathit{idf}}
\def\VSMScore{\mathrm{VSMScore}}
\def\rVSMScore{\mathrm{rVSMScore}}
\def\SimiScore{\mathrm{SimiScore}}
\def\BugLocator{\mathrm{BugLocator}}
\def\Success#1{Success${}_{@#1}$\xspace}

\def\Pone{$P_1$\xspace}
\def\Poned{$P_1^d$\xspace}
\def\Ponei{$P_1^i$\xspace}
\def\Ptwo{$P_2$\xspace}

\newcommand{\BLT}[1]{#1\xspace}

\begin{document}

\title{Evaluation of Cross-Lingual Bug Localization:\\Two Industrial Cases}

\author{%
  \IEEEauthorblockN{Shinpei Hayashi}
  \IEEEauthorblockA{%
  \textit{School of Computing}\\
  \textit{Tokyo Institute of Technology}\\
  Tokyo 152--8550, Japan \\
  hayashi@c.titech.ac.jp}
  \and
  \IEEEauthorblockN{Takashi Kobayashi}
  \IEEEauthorblockA{%
  \textit{School of Computing}\\
  \textit{Tokyo Institute of Technology}\\
  Tokyo 152--8550, Japan \\
  tkobaya@c.titech.ac.jp}
  \and
  \IEEEauthorblockN{Tadahisa Kato}
  \IEEEauthorblockA{%
  \textit{Research \& Development Group}\\
  \textit{Center for Digital Services, Hitachi, Ltd.}\\
  Kanagawa 244--0817, Japan \\
  tadahisa.kato.en@hitachi.com}
}

\maketitle
\thispagestyle{plain}

\begin{abstract}
  This study reports the results of applying the cross-lingual bug localization approach proposed by Xia \etal\ to industrial software projects.
  To realize cross-lingual bug localization, we applied machine translation to non-English descriptions in the source code and bug reports, unifying them into English-based texts, to which an existing English-based bug localization technique was applied.
  In addition, a prototype tool based on \BLT{BugLocator} was implemented and applied to two Japanese industrial projects, which resulted in a slightly different performance from that of Xia \etal
\end{abstract}

%%%%%%%%%%%%%%%%%%%%%%%%%%%%%%%%%%%%%%%%%%%%%%%%

\section{Introduction}\label{s:introduction}

In software development, fixing bugs is time consuming.
It is fundamentally difficult for developers to keep writing bug-free source code consistently, as bugs frequently occur.
Bugs found by users or other developers are reported in bug reports and reviewed by developers so that they can be fixed~\cite{anvik2005coping}.
When fixing a bug by referring to a bug report, the modules that need to be fixed must be identified to resolve the bug.
This process, known as bug localization~(BL), is a time-consuming and labor-intensive task in larger projects~\cite{planning2002economic}.

Automated BL techniques have been studied to improve the bug-fixing efficiency.
There are several automated BL approaches, including dynamic analysis, static analysis, history analysis, and information retrieval~(IR).
Among these, all except for the IR-based approach require additional information, such as execution scenarios and a precise analysis of the program.
The dynamic analysis approach includes fault localization techniques that record and analyze the test execution characteristics and test results for each case, thereby suggesting causes for the bugs.
Unlike open-source software, users are not expected to perform an initial analysis on the conditions to reproduce defects or narrow down their causes.
In many cases, a bug report is the only primary information on defects.
In this study, the easiest IR approach that does not require additional information was used.

Most existing IR-based BL techniques assume that bug reports are written in English, whereas source files are written using English-based identifiers~\cite{buglocator,saha2013improving,wong2014boosting}.
This is because identifiers in source code are considered a sequence of English terms such that the focus is on their lexical similarity to the English descriptions in bug reports.
However, native non-English developers can also write bug reports in languages other than English, particularly for projects involving stakeholders from non-English-speaking countries.
In such cases, the use of a non-English language is motivated to facilitate communication among developers.
However, developers who receive non-English bug reports face difficulty in using existing BL approaches, which are designed for English text.

To mitigate this gap, Xia \etal\ proposed \BLT{CrosLocator}, a cross-lingual BL technique.
It allows bug localization in reports written in languages other than English by applying machine translation.
In particular, they examined the BL performance for Chinese bug reports~\cite{croslocator}.
Compared to the performance in the study of \BLT{BugLocator} evaluation~\cite{buglocator}, \BLT{CrosLocator}'s performance was insufficient as a BL method after machine translation.
Xia \etal\ concluded that cross-lingual IR is more challenging than normal IR, and cross-lingual BL is even more challenging than normal BL\@.
However, to date, there have been few reported applications of cross-lingual BL\@.
To the best of our knowledge, only the study by Xia \etal\ has been reported as applying the approach, and it focused solely on a Chinese open-source project.

This paper reports the results of a feasibility study on cross-lingual BL for two industrial projects in Japan.
The main contributions of this study are as follows:
\begin{itemize}
  \item Improving the performance of the cross-lingual BL approach by applying machine translation for source code, which includes Japanese texts, along with bug reports.
  \item Investigating the performance of two industrial projects.
\end{itemize}

The remainder of this paper is organized as follows:
\Cref{s:preliminary} explains IR-based BL approaches and the baseline cross-lingual BL approach briefly.
\Cref{s:technique} describes the extended cross-lingual BL approach used in this study.
\Cref{s:exp} presents an evaluation of the approach using two industrial cases.
Finally, \cref{s:conclusion} concludes the paper and outlines future research.

%%%%%%%%%%%%%%%%%%%%%%%%%%%%%%%%%%%%%%%%%%%%%%%%

\section{Preliminaries}\label{s:preliminary}

\subsection{IR-Based Bug Localization}
In IR-based BL, source files and bug reports are considered documents, and their similarity is calculated using text analysis.
IR-based BL calculates the similarity between a given bug report and each source code, and outputs a ranked list of source files based on the calculated similarity.

\subsubsection{VSM}
The vector space model~(\BLT{VSM}) is an IR approach that calculates the similarity between documents by vectorizing them.
Term frequency-inverse document frequency~(\textit{tf-idf}) is typically used in vectorization, which considers the importance of terms in the documents.

Among various tf-idf computation methods, we adopted the following definition proposed by Zhou \etal~\cite{buglocator}:
\begin{align*}
  \tf(t,d) &= \log \left( \frac{c_{t,d}}{c_d} + 1 \right), \\
  \idf(t)  &= \log \frac{|D|}{|\{\, d \mid c_{t,d} > 0 \,\}|}
\end{align*}
where $d \in D$ denotes a document; $t \in T$ denotes a term; $c_{t,d}$ is the number of occurrences of the term $t$ in document $d$; and $c_d$ is the total number of terms in $d$.
Term frequency $\tf$ increases as the number of occurrences of the term increases.
The inverse document frequency $\idf$ decreases as the term appears in more documents.
Assuming that terms with high generality are less important, the product $\tf(t,d) \cdot \idf(t)$ is used for the weight of $t$ in $d$.
Document $d$ is vectorized as a $|T|$-dimensional vector $\vec{d}$ whose elements are the weights of each term.

The similarity score between a given source code $d$ and a bug report $q$ is calculated as the cosine similarity of the vectors:
\begin{align*}
   \cos(d_1, d_2) &= \frac{\vec{d_1} \cdot \vec{d_2}}{|\vec{d_1}| |\vec{d_2}|}, \\
  \VSMScore(q, d) &= \cos(q, d).
\end{align*}

Large source files tend to have a high probability of containing bugs.
However, in simple \BLT{VSM}, the score of a large source file tends to be lower than expected.

\subsubsection{rVSM}
The revised VSM~(\BLT{rVSM})~\cite{buglocator} is a \BLT{VSM} that addresses the aforementioned issue by taking into account the number of terms in a file:
\begin{align*}
             N(c) &= \frac{c - c_{\min}}{c_{\max} - c_{\min}}, \\
  \rVSMScore(q,d) &= \frac{1}{1 + e^{-N(c_d)}} \VSMScore(q,d)
\end{align*}
where $N(c)$ is the linear normalization of the number of terms $c$ to $[0, 1]$; $c_{\max}$ and $c_{\min}$ are the maximum and minimum number of terms in the target document set, respectively.
The larger the number of terms, \ie, the larger the file size, the closer $N(c)$ is to 1, and the correction weight of $\VSMScore$ becomes larger.
This means that $\rVSMScore(q,d)$ produces a higher value as the source code document $d$ increases in size, thereby considering file size when calculating the similarity.

\subsubsection{BugLocator}
\BLT{BugLocator}~\cite{buglocator}, extended from \BLT{rVSM}, considers previous bug reports that have already been resolved.
\begin{align*}
   \SimiScore(q,d) &= \sum_{b \in B_d} \frac{\cos(q,b)}{n_b}, \\
  \BugLocator(q,d) &= (1 - \alpha) \, N(\rVSMScore(q,d)) + {} \\
                   &\hspace{1.24cm} \alpha \, N(\SimiScore(q,d))
\end{align*}
where $B_d$ is the set of previous bug reports on source files that need to be fixed, and $N_b$ is the number of fixed files required to resolve bug $b$.
Parameter $\alpha \in [0, 1]$ determines the ratio at which $\rVSMScore$ and $\SimiScore$ are combined.
The case of $\alpha = 0$ is equivalent to \BLT{rVSM} and does not consider any previous bug information.
When $\alpha = 1$, only the similarity with past bug reports is considered, and $\BugLocator$ increases when the target bug is similar to past bugs.

\subsection{Cross-Lingual Bug Localization}\label{s:clir}

Cross-lingual (or cross-language) IR retrieves documents that are written in a language different from the input query~\cite{kishida2005technical}.
Several cross-lingual IR approaches have been proposed.
Hayes \etal\ used Google Translate to apply machine translation to Italian sentences to obtain English sentences and recover software traceability~\cite{hayes2011software}.
Xu \etal\ performed corpus-based translation on a corpus that was pre-created by crawling websites and querying input texts, thereby using Chinese search queries on English websites to enable question retrieval~\cite{xu2018domain}.
Tonoike \etal\ extended term dictionaries by creating a corpus of compound nouns non-existent in dictionaries, thus improving the performance in Japanese--English translation estimation~\cite{tonoike2006comparative}.

\begin{figure}[tb]\centering
  \includegraphics[width=\linewidth]{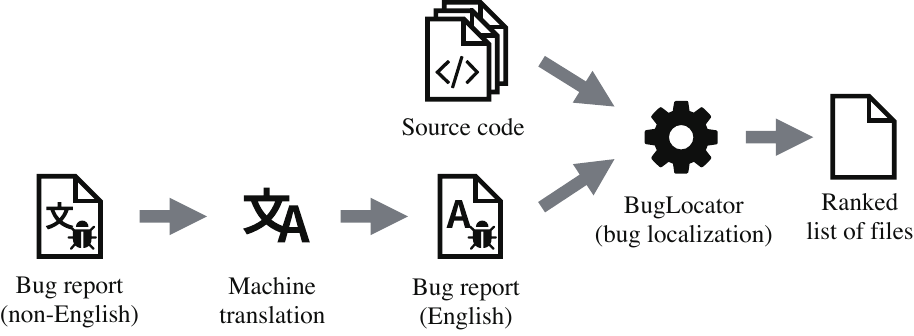}
  \caption{Overview of cross-lingual bug localization via \BLT{CrosLocator}.}\label{f:crosslang-bl}
\end{figure}
Xia \etal\ proposed \BLT{CrosLocator}, a cross-lingual BL technique applied to bug reports written in a non-English language by applying machine translation to the non-English bug report input.
An overview of cross-lingual BL using \BLT{CrosLocator} is presented in \cref{f:crosslang-bl}.
The input consists of source code based on English and a bug report written in a non-English language.
First, machine translation is applied to the natural language descriptions in the bug report for translation into English.
Next, \BLT{BugLocator}, an IR-based BL technique is run, with the source code and bug report translated into English as inputs, thereby obtaining a ranked list of source files that are likely to require fixing.
\BLT{BugLocator} utilizes bug reports similar to the given bug report to find relevant source files based on the textual similarity between the bug reports and source files.
Note that this figure only shows one machine translator, whereas the original \BLT{CrosLocator} uses multiple translators~(see \cref{s:technique} for details).

\section{Extension for \BLT{CrosLocator}}\label{s:technique}

We extended \BLT{CrosLocator} to perform cross-lingual BL\@.
We analyzed several software projects developed in Japan and found that comments and string literals in the source code sometimes included Japanese descriptions as well as bug reports.
As Karampatsis \etal~\cite{karampatsis-icse2020} reported, non-English terms can be found in strings and comments in source code due to the developer's primary language or internationalization purposes.
Such Japanese terms in string literals or comments are processed directly as regular inputs by \BLT{BugLocator} in the later stage.
In contrast, the Japanese descriptions in the given bug report were translated into English and input into \BLT{BugLocator}; therefore, even if both the bug report and source code contained similar Japanese descriptions, they did not contribute to the similarity degree.

\begin{figure}[tb]\centering
  \includegraphics[width=\linewidth]{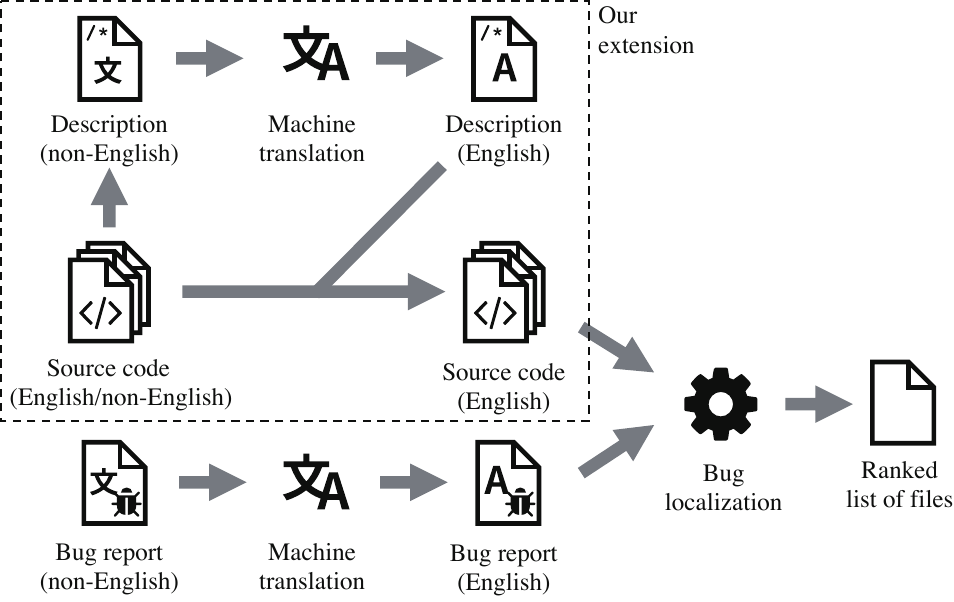}
  \caption{Extension for \BLT{CrosLocator}.}\label{f:crosslang-bl-ex}
\end{figure}
Based on this observation, the similarity between each description was determined by unifying the descriptions into English in advance.
An overview of the \BLT{CrosLocator} approach is shown in \cref{f:crosslang-bl-ex} with an extension for cases in which Japanese descriptions are also included in the source code.
The source code was parsed to extract string literals and comment descriptions.
Among the extracted descriptions, those containing at least one Japanese character were considered Japanese descriptions.
Machine translation was then applied to translate them into English.
Japanese characters were detected based on the range of character codepoints.
The translated English descriptions were embedded in the same position from which the source code was extracted, replacing the original English descriptions.
The entire source code was not translated directly because such an automated translation may break the identifiers and syntax of the source code.
If a comment contained expressions corresponding to Japanese literals, only the literals were extracted and translated because translating the entire comment would lead to the same issue.
Although this extension is simple, it can provide significant accuracy gains at low cost in cross-language BL when there are non-English terms in the source code.

A single machine translator was used for the \BLT{CrosLocator} approach.
The original \BLT{CrosLocator} uses multiple localization results with multiple English texts obtained from multiple machine translators and computes the score by merging the multiple results through the data fusion technique~\cite{data-fusion}.
However, sometimes it is not easy to have multiple translators available.
Sending bug reports and source code to many external translation services may be impractical due to confidentiality concerns, especially in corporate use.
Therefore, it was assumed that only a limited number of reliable machine translators are available for BL\@.
For both confidentiality and practical use, only the cloud-based neural machine translation service contracted by the authors' organization was available for machine translation to be used in this study because the confidential software assets of the organization were to be shared.

The \BLT{BugLocator} implementation bundled into Bench4BL~\cite{bench4bl} was used to implement the extended \BLT{CrosLocator}.
We revised the original implementation, which only accepted Java programs as input source code, to allow inputs from other programming languages.

%%%%%%%%%%%%%%%%%%%%%%%%%%%%%%%%%%%%%%%%%%%%%%%%

\begin{table}[tb]\centering
  \caption{Target Projects}\label{t:projects}
  \begin{tabular}{llrr} \hline
             & Main                 & \# source & \# bug \\
    Projects & programming language &     files & reports \\ \hline
    \Pone & C\#  & 929 &   8 / 186 \\
    \Ptwo & Java & 591 & 267 / 316 \\ \hline
  \end{tabular}
\end{table}

\section{Experiment}\label{s:exp}

In this experiment, we aim to answer two research questions~(RQs):
\def\RQone{Is the cross-lingual bug localization approach effective for industry projects?}
\def\RQtwo{Does translating texts in source files contribute to the improvement of bug localization performance?}
\begin{itemize}
  \item \textbf{\RQ{1} (Replication of \BLT{CrosLocator} approach):} \RQone
  \item \textbf{\RQ{2} (Effectiveness of the extension):} \RQtwo
\end{itemize}

\begin{table*}[tb]\centering
  \caption{Experimental Results}\label{t:results}
  {\begin{tabular}{lclrcccc} \hline
    Project & Extension${}^\dagger$ & Base technique & MAP & MRR & \Success{5} & \Success{10} \\ \hline
    \multirow{4}{*}{\Poned}
      & Yes & \BLT{BugLocator} & 0.224 & 0.322 & 0.375 & 0.500 \\
      & Yes & \BLT{rVSM}       & 0.158 & 0.248 & 0.375 & 0.375 \\
      & No  & \BLT{BugLocator} & 0.107 & 0.201 & 0.250 & 0.375 \\
      & No  & \BLT{rVSM}       & 0.085 & 0.175 & 0.125 & 0.250 \\ \hline
    \multirow{4}{*}{\Ponei}
      & Yes & \BLT{BugLocator} & 0.157 & 0.276 & 0.375 & 0.625 \\
      & Yes & \BLT{rVSM}       & 0.145 & 0.265 & 0.375 & 0.500 \\
      & No  & \BLT{BugLocator} & 0.112 & 0.139 & 0.250 & 0.375 \\
      & No  & \BLT{rVSM}       & 0.097 & 0.178 & 0.125 & 0.250 \\ \hline
    \multirow{2}{*}{\Ptwo}
      &(No) & \BLT{BugLocator} & 0.174 & 0.270 & 0.390 & 0.461 \\
      &(No) & \BLT{rVSM}       & 0.145 & 0.220 & 0.326 & 0.397 \\ \hline
  \end{tabular}}\\
  $\dagger$ Extension in \cref{s:technique}: translating non-English texts in source files
\end{table*}

\subsection{Experimental Setup}

\subsubsection{Dataset}

The projects used in this experiment are presented in \cref{t:projects}.
\Pone is a software product written mainly in C\# and developed by a company.
\Ptwo is written in Java and developed by the same company.

The bug reports maintained for each project were used in the study.
Because an interview with a developer of the product in \Pone was conducted, as described below, the number of bug reports had to be limited to a small number.
We filtered all 186 recorded bug reports based on the following criteria:
\begin{enumerate}
  \item excluding bug reports unrelated to functional bugs (retaining 13 out of 186),
  \item excluding bug reports for which bug fixes had not been completed at the latest code snapshot (retaining 9 out of 13), and
  \item excluding a bug report in which a C\# source file was not fixed when resolving the bug (retaining 8 out of 9),
\end{enumerate}
thereby obtaining eight bug reports for \Pone.
In the case of \Ptwo, 267 out of the 316 bug reports met the criterion of fixing at least one Java source file when resolving the bugs and were used for this study.

These projects differ in the preparation of oracle lists of files related to the bug reports, \ie, the ground truth files that should be identified by a BL technique from the given bug reports.
In \Ptwo, oracle files were automatically identified based on issue-commit linking.
Bug reports and source code were managed using issue tracking and version control systems, respectively.
The issue tracking system recorded in which commit the bug was resolved.
The source files that were modified in the corresponding commit were identified as the files that needed to be fixed to resolve the bug.
By associating these files with the bug report, the source files were determined as the ideal BL results corresponding to the bug report.
However, this automated approach does not always identify the correct solution in a bug report.
For example, incorrect or missing correspondences between bugs and fixing commits recorded in the issue tracking system or tangled commits~\cite{tangled-changes} that include other changes along with the fixing of a bug can lead to an incorrectly deduced automatic solution.
Additionally, in the BL process, it is practical to recommend files to developers that enhance their understanding of the bug, even if those files do not need to be fixed when resolving the bug.
If only fixed files are considered correct, an evaluation from this perspective cannot be performed.

However, in \Pone, we identified a mapping between bug reports and solution files based on an interview with an actual developer of \Pone.
The following two file types were identified for the eight bug reports:
\begin{itemize}
  \item $O^d$: files containing \emph{direct} code fragments to be fixed when resolving the bug, and
  \item $O^i$: files \emph{indirectly} related and contributing to bug identification, but not containing direct code fragments.
\end{itemize}
To determine if the cross-lingual BL approach can identify not only direct code fragments but also indirect ones, we prepared an experimental treatment that included indirectly related files as oracles (\Ponei), in addition to another treatment that targeted only files containing direct bug locations as oracles (\Poned).
In the case of \Ponei, $O = O^d \cup O^i$ was used for the oracles.

The effect of our extension to translate non-English texts into source files introduced in \cref{s:technique} was evaluated using only \Pone because the source code of \Ptwo does not include any Japanese descriptions.

\subsubsection{Evaluation Criteria}

Mean average precision~(MAP)~\cite{map}, mean reciprocal rank~(MRR)~\cite{mrr}, and \Success{N} (percentage of rankings with at least one oracle file within the top $N$) were used as metrics to evaluate BL and other ranking-based tasks.
These metrics were measured using trec\_eval~\cite{trec-eval}.
We used $N=5$ and $N=10$ for \Success{N}, based on evidence of practitioners' expectations of BL\cite{practitioners-expectations}.

%%%%%%%%%%%%%%%%%%%%%%%%%%%%%%%%%%%%%%%%%%%%%%%%

\subsection{\RQ{1}: \RQone}

The extended \BLT{CrosLocator} was applied to \Poned, \Ponei, and \Ptwo.
The results are presented in \cref{t:results}.

\Heading{Overview}
When \BLT{BugLocator} was used as the base bug localizer, \Success{5} for \Poned was 0.375, indicating that at least one oracle file was recommended within the top five entries in 37.5\% of bug reports.
In addition, \Success{10} was 0.500, meaning that half of the results recommended at least one oracle file within the top 10 entries, highlighting the usefulness of the results.
The MAP results for \Ponei were generally lower than those for \Poned, indicating the difficulty in identifying more buggy files, including indirect oracles.
The results using \BLT{BugLocator} for \Poned produced better MRR scores than those for \Ponei, suggesting that \BLT{BugLocator}'s feature of utilizing the relevant source files of similar bug reports may have had a negative effect when using indirect oracles.

\Heading{Confirmation of the effectiveness of \BLT{BugLocator} compared to rVSM}
Overall, the accuracy improved when \BLT{BugLocator} was used instead of \BLT{rVSM}.
In other words, the performance improvement when using similar bug reports was effective as in \BLT{BugLocator}.
The results for \Ptwo were similar to those of \Pone.

\Heading{Comparison between cross-lingual and non-cross-lingual settings}
Xia \etal~\cite{croslocator} reported that cross-lingual BL is challenging because the accuracy of the \BLT{CrosLocator} results (MAP of 0.116 for Ruby-China) was inferior to the accuracy reported in the paper proposing \BLT{BugLocator} for English bug reports (MAP of 0.22--0.545 for Zwing, SWT, AspectJ, and Eclipse)~\cite{buglocator}, although the target projects were different.

Although the values obtained in this experiment (MAP of 0.157 and 0.224 for \Poned, \Ponei, and \Ptwo) did not reach the ideal levels of \BLT{BugLocator}, they represented an improvement over the values reported for the extended \BLT{CrosLocator}.
Considering the other evaluation metrics mentioned above, we believe that this approach is applicable to BL in Japanese.

\Conclusion{The cross-lingual BL with text translation of source files achieved reasonable performance.}

%%%%%%%%%%%%%%%%%%%%%%%%%%%%%%%%%%%%%%%%%%%%%%%%

\subsection{\RQ{2}: \RQtwo}

\begin{figure}[tb]\centering
  \includegraphics[width=8.5cm]{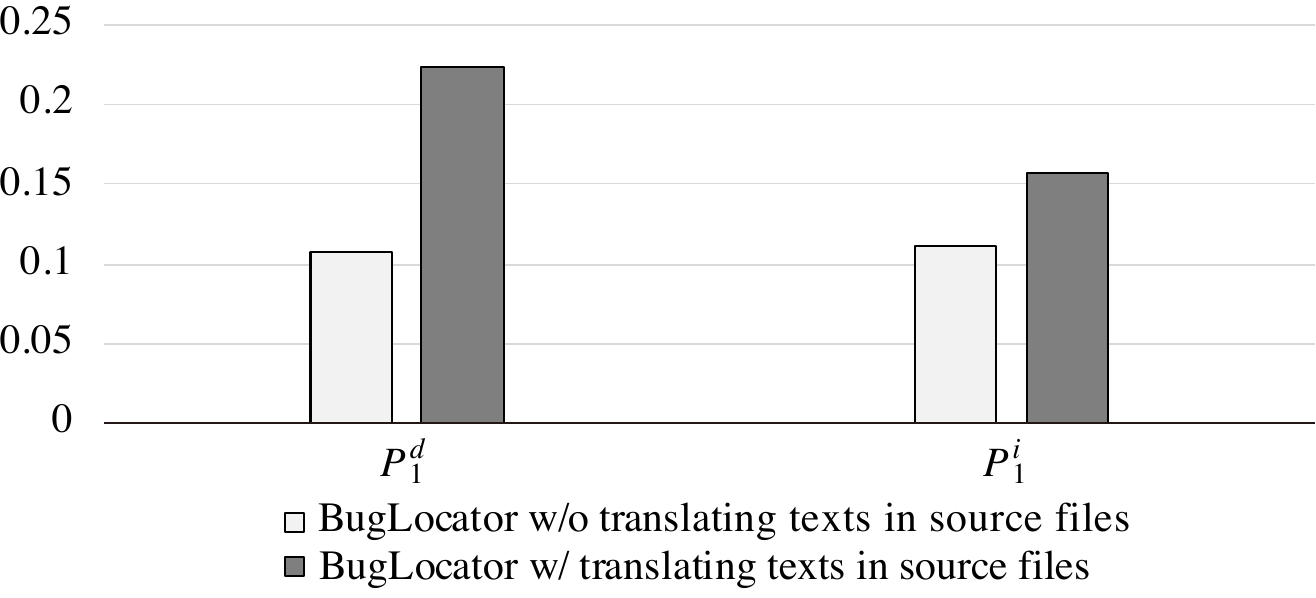}
  \caption{MAP performance with and without text translation of source files.}\label{f:translation-result}
\end{figure}
The extended \BLT{CrosLocator}, with text translation of source files, was applied to the projects.
The ``Extension'' column of \cref{t:results} indicates whether translation was applied to the texts in the source files for the technique in \cref{s:technique}.
\Cref{f:translation-result} provides a summary of the improvement in MAP performance when the text translation feature is available.
When applying \BLT{BugLocator}, all evaluation metrics showed significant improvement when text translation was applied.

In a bug report of \Pone, the ranking of the oracle file improved from 91st to 11th when text translation was applied.
Japanese domain-specific terms corresponding to the terms in the bug report were included in the comments in the file, and their translation led to better lexical matching, resulting in an improvement in the ranking.

\Conclusion{The translation of texts in source files was effective in improving localization performance.}

\section{Conclusion}\label{s:conclusion}

This study discusses the applicability of cross-lingual BL to Japanese descriptions.
We applied the method to two industrial projects and achieved satisfactory results.
To enhance the effectiveness of the BL approach, we plan to integrate a third-generation BL approach as the base bug localizer, thereby improving matching efficiency~\cite{bugzbook}.
Moreover, we will extend the application of the cross-lingual BL approach to languages other than Chinese and Japanese.
Lastly, we intend to apply this approach to additional projects, collecting cases that are based on the real localization settings employed by practitioners.

\newpage
%\bibliographystyle{IEEEtran}
%\bibliography{references}

% Generated by IEEEtran.bst, version: 1.12 (2007/01/11)

\end{document}